\documentclass[11pt]{eptcs}
\pdfoutput=1

\usepackage{amsfonts,amsmath} 
\usepackage{tikz}
\usetikzlibrary{arrows,calc,decorations.pathmorphing,decorations.markings,backgrounds,decorations,decorations.pathmorphing,positioning,fit,automata,shapes,patterns,decorations.text}

\usepackage{tabulary}
\newcolumntype{K}[1]{>{\centering\arraybackslash}p{#1}}

\usepackage{outlines}

\usepackage{enumitem}
\setenumerate[1]{label=\Roman*.}
\setenumerate[2]{label=\Alph*.}
\setenumerate[3]{label=\roman*.}
\setenumerate[4]{label=\alph*.}

\usepackage{graphicx} 
\usepackage{color}  
\usepackage{xspace}
\usepackage{subfig}

\newcommand{\toolname}{Sim-ATAV\xspace}
\newcommand{\GUR}{Global UR\xspace}
\newcommand{\CAUR}{CA+UR\xspace}
\newcommand{\CASA}{CA+SA\xspace}

\newcommand{\matlab}{MATLAB\textsuperscript{\circledR}\xspace}

\newcommand{\reals}{\ensuremath{\mathbb{R}}}
\renewcommand{\int}{\ensuremath{\mathbb{Z}}}
\newcommand{\model}{\ensuremath{M}}
\newcommand{\modeld}{\ensuremath{M_d}}
\newcommand{\phid}{\ensuremath{\phi_d}}
\newcommand{\X}{\ensuremath{\mathcal{X}}}
\newcommand{\x}{\ensuremath{{\bf x}}}
\newcommand{\y}{\ensuremath{{\bf y}}}
\newcommand{\U}{\ensuremath{\mathcal{U}}}
\renewcommand{\u}{\ensuremath{{\bf u}}}
\renewcommand{\P}{\ensuremath{\mathcal{P}}}
\newcommand{\p}{\ensuremath{{\bf p}}}
\newcommand{\Pc}{\ensuremath{\mathcal{W}}}
\newcommand{\pc}{\ensuremath{{\bf w}}}
\newcommand{\Pd}{\ensuremath{\mathcal{V}}}
\newcommand{\pd}{\ensuremath{{\bf v}}}

\renewcommand{\sim}{\ensuremath{sim}}

\newcommand{\statetrace}{\ensuremath{X}}
\newcommand{\inputtrace}{\ensuremath{U}}

\newcommand{\Uc}{\mathcal{U}}
\newcommand{\Ic}{\mathcal{I}}
\newcommand{\Oc}{\mathcal{O}}
\newcommand{\dle}{[\![} 
\newcommand{\dri}{]\!]} 
\newcommand \genMet {\mathbf{d}}
 
\newcommand \robsemD[1] {\dle #1 \dri_{\genMet}} 
\newcommand{\tss}{T}
 
\newcommand \copreals {\reals_{\geq 0}}

\newtheorem{definition}{Definition}
\newtheorem{exmp}{Example}[section]

\providecommand{\publicationstatus}{Appeared in The 2018 29\textsuperscript{th} IEEE Intelligent Vehicles Symposium (IV 2018)}

\title{\LARGE \bf
Simulation-based Adversarial Test Generation for Autonomous Vehicles with Machine Learning Components
\thanks{This work was partially funded by NSF awards CNS 1446730, 1350420}
\thanks{Source Code is available at https://cpslab.assembla.com/spaces/sim-atav}
}

\author{Cumhur Erkan Tuncali\textsuperscript{1} \quad\ Georgios Fainekos\textsuperscript{2} \quad\ Hisahiro Ito\textsuperscript{1} \quad\ James Kapinski\textsuperscript{1}
		\institute{\textsuperscript{1}Toyota Research Institute of North America, Ann Arbor, MI, USA\\
		\textsuperscript{2} Arizona State University, Tempe, AZ, USA}
	\email{cumhur.tuncali@toyota.com, fainekos@asu.edu, \{hisahiro.ito, jim.kapinski\}@toyota.com}}
	
\def\titlerunning{Simulation-based Adversarial Test Generation for Autonomous Vehicles with ML Components}
\def\authorrunning{C.E.~Tuncali, G.~Fainekos, H.~Ito \& J.~Kapinski}

\begin{document}

\maketitle

\begin{abstract}
Many organizations are developing autonomous driving systems, which are expected to be deployed at a large scale in the near future. Despite this, there is a lack of agreement on appropriate methods to test, debug, and certify the performance of these systems. One of the main challenges is that many autonomous driving systems have machine learning components, such as deep neural networks, for which formal properties are difficult to characterize. We present a testing framework that is compatible with test case generation and automatic falsification methods, which are used to evaluate cyber-physical systems. We demonstrate how the framework can be used to evaluate closed-loop properties of an autonomous driving system model that includes the ML components, all within a virtual environment. We demonstrate how to use test case generation methods, such as covering arrays, as well as requirement falsification methods to automatically identify problematic test scenarios. The resulting framework can be used to increase the reliability of autonomous driving systems.
\end{abstract}

\section{Introduction} \label{sec:intro}

Many groups are developing autonomous driving systems.
These systems are on the road now and are expected to have a significant impact on the vehicle market and the broader economy in the near future; however, no generally agreed upon testing or verification methods have arisen for these systems. One reason for this is that the current designs usually include some machine learning (ML) components, such as deep neural networks (DNNs), which are notoriously difficult to test and verify. We present a framework for Simulation-based Adversarial Testing of Autonomous Vehicles (\toolname),  which can be used to check closed-loop properties of autonomous driving systems that include ML components. We describe a testing methodology, based on a test case generation method, called covering arrays, and requirement falsification methods to automatically identify problematic test scenarios. The resulting framework can be used to increase the reliability of autonomous driving systems.

Autonomous driving system designs often use ML components such as DNNs to classify objects within CCD images and to determine their positions relative to the vehicle, a process known as \emph{semantic segmentation} \cite{geiger2012we}. Other designs use neural networks (NNs) to perform \emph{end-to-end} control of the vehicle, meaning that the NN takes in the image data and outputs actuator commands, without explicitly performing an intermediate step to do the semantic segmentation \cite{pomerleau1989alvinn}. Still other approaches use end-to-end learning to do intermediate decisions like risk assessment \cite{strickland2017deep}. 

The ML system components are problematic from an analysis perspective, as it is difficult or impossible to characterize all of the behaviors of these components under all circumstances. One reason for this is that the complexity of these systems, in terms of number of parameters, can be high. For example, AlexNet, a pre-trained, DNN, which is used for classification of CCD images, has 60 million parameters \cite{krizhevsky2012imagenet}. Another reason for the difficulty in characterizing behaviors of the ML components is that the parameters are learned based on training data. Characterizing the ML behaviors is, in some ways, as difficult as the task of characterizing the training data. Again using the AlexNet example, the number of training images used was 1.2 million. While a strength of DNNs is their ability to generalize from training data; the challenge for analysis is that we do not well understand how they generalize for all possible cases.

There has been significant interest recently on verification and testing for ML components (see Sec.~\ref{sec:related}). For example, adversarial testing approaches seek to identify perturbations in image data that result in misclassifications. By contrast, our work focuses on methods to determine perturbations in the configuration of a testing scenario, meaning that we seek to find scenarios that lead to unexpected behaviors, such as misclassifications and ultimately collisions. The framework that we present allows this type of testing in a virtual environment.
By utilizing advanced 3D models and image rendering tools, such as the ones used in game engines or film studios, the gap between testing in a virtual environment and the real world can be minimized.

Most of the previous work to test and verify systems with ML components focuses only on the ML components themselves, without consideration of the closed-loop behavior of the system. For autonomous driving applications, we observe that the ultimate goal is to evaluate the closed-loop performance, and so the testing methods used to evaluate these systems should reflect this.

The closed-loop nature of a typical autonomous driving system can be described as follows. A \emph{perception} system processes data gathered from various sensing devices, such as cameras, LIDAR, and radar. The output of the perception system is an estimation of the principal (\emph{ego}) vehicle's position with respect to external obstacles (e.g., other vehicles, called \emph{agent} vehicles, and pedestrians). A \emph{path planning} algorithm uses the output of the perception system to produce a short-term plan for how the ego vehicle should behave. A \emph{tracking controller} then takes the output of the path planner and produces actuation outputs, such as accelerator, braking, and steering commands. The actuation commands affect the vehicle's interaction with the environment. The iterative process of sensing, processing, and actuating is what we refer to as closed-loop behavior.

Our contributions can be summarized as follows. 
We provide a new algorithm to perform falsification of formal requirements for an autonomous vehicle in a closed-loop with the perception system, which includes an efficient means of searching over discrete and continuous parameter spaces. Our method represents a new way to do adversarial testing in scenario \emph{configuration space}, as opposed to the usual method, which considers adversaries in \emph{image space}.
Additionally, we demonstrate a new way to characterize problems with perception systems in \emph{configuration space}. Lastly, we extend software testing notions of \emph{covering arrays} to closed-loop cyber-physical system (CPS) applications based on ML.

\section{Related work} \label{sec:related}

Verification and testing for NNs is notoriously difficult since NNs correspond to complex nonlinear and non-convex functions.
Currently, the methods developed for the analysis of NN components can be classified into two main categories.

The first category concerns output range verification for a given bounded set of input values \cite{DuttaEtAl2017arxiv,HuangEtAl2017cav,KatzEtAl2017cav,PulinaT2010cav}.
The methods in this category use a combination of Satisfiability Modulo Theory (SMT) solvers, Linear Programing (LP), gradient-based local search, and Mixed Integer Linear Programming (MILP).
The second category deals with adversarial sample generation (\cite{PeiEtAl2017sosp,TianEtAl2017arxiv,PapernotEtAl2016eurosp,PapernotEtAl2017asiaccs,GoodfellowSS2015iclr,ChenEtAl2017arxiv}), 
which addresses the problem of how to minimally perturb the input to the NN so that the classification decision of the NN changes.
If these perturbations are very small, then in a sense, the NN is not robust.
The methods typically employed to solve this problem primarily use some form of gradient ascent to adjust the input so that the output of the NN changes. 
The differences between the various approaches relate to whether the system is black-box or white-box and how the gradient is computed or approximated.
Under this category, we could also potentially include generative adversarial networks \cite{Goodfellow17tutorial}.

All of the aforementioned methods deal with the verification and testing of NNs at the component level; however, our work targets the NN testing problem at the system level.
The line of research that is the closest in spirit to our work is \cite{DreossiEtAl2017rmlw,DreossiDS2017nfm,Dreossi2018}.
The procedure described in \cite{DreossiEtAl2017rmlw,DreossiDS2017nfm,Dreossi2018} analyzes the performance of the perception system using static images to identify candidate counterexamples, which they then check using simulations of the closed-loop behaviors to determine whether the system exhibits unsafe behaviors.
%
%
We, on the other hand, provide a new method to search for unsafe behaviors directly on the closed-loop behaviors of the system, meaning our search uses a global optimizer guided by a cost function that is defined based on the closed-loop behaviors.

The goals of our approach are similar to those of \cite{OKelly2017}, though their framework is not capable of simulating the autonomous vehicle system in a closed-loop, including the perception system.
In contrast to our work, their framework utilize a game engine only to visualize the results after the testing and analysis is done on well defined, but simpler, vehicle dynamic models like a bicycle model, which may not well represent real behaviors.

Beyond NN-focused testing and analysis, our work borrows ideas from robustness guided falsification for autonomous vehicles \cite{TuncaliPF16itsc}, where the goal is to detect boundary-case failures.
Finally, our work falls under the broader context of automatic test generation methods for autonomous vehicles and driver assist systems \cite{KimKDS17esl,KimJSSY16emsoft}, but our methods and testing goals are different.

\section{Preliminaries} \label{sec:preliminaries}

This section presents the setting used to describe the testing procedures performed using our framework. The purpose of our framework is to provide a mechanism to test, evaluate, and improve on an autonomous driving system design. To do this, we use a simulation environment that incorporates models of a vehicle (called the \emph{ego} vehicle), a perception system, which is used to estimate the state of the vehicle with respect to other objects in its environment, a controller, which makes decisions about how the vehicle will behave, and the environment in which the ego vehicle exists. The environment model contains representations of a wide variety of objects that can interact with the ego vehicle, including roads, buildings, pedestrians, and other vehicles (called \emph{agent} vehicles). The behaviors of the system are determined by the evolution of the model states over time, which we compute using a \emph{simulator}. 

Formally, the framework implements a model of the system, which is a tuple \[\model=(\X ,\U, \P ,\sim ),\] where $\X$ is a set of system states, $\U$ is a set of inputs, and $\sim$ is a simulation function \[\sim :\X \times \U \times \P \times T \rightarrow \X,\] where $T$ is a discrete set of sample times $t_0, t_1, \ldots, t_N$, with $t_i<t_{i+1}$. $\P=\Pc \times \Pd$ is a combination of continuous-valued and discrete-valued parameters, where $\Pc=\Pc_1\times \cdots \times \Pc_W$ and each $\Pc_i \subseteq \reals$, and $\Pd=\Pd_1\times \cdots \times \Pd_V$ and each $\Pd_i$ is some finite domain, such as Boolean or a finite list of agent car colors.

Given $\x \in \X$, $\hat{\x}=\sim(\x,\u,\p,t)$ is the state reached starting from state $\x$ after time $t\in T$ under input~$\u \in \U$ and parameter value $\p\in \P$. We call a sequence \[\inputtrace=(\u_0,t_0)(\u_1,t_1)\cdots (\u_N,t_N),\] where each $\u_i\in \U$ and $t_i\in T$, an input trace of $\model$. Given a model $\model$, an input trace of $\model$, $\inputtrace$, and a~$\p\in \P$, a simulation trace of $\model$ under input $\inputtrace$ and parameters $\p$ is a sequence \[T=(\x_0,\u_0,t_0)(\x_1,\u_1,t_1)\cdots (\x_N,\u_N,t_N),\] where $\sim(\x_{i-1},\u_{i-1},\p,t_{i-1})=\x_i$ for each $1\leq i\leq N$. For a given simulation trace $T$, we call $\statetrace=(\x_0,t_0)(\x_1,t_1)\cdots (\x_N,t_N)$ the state trace. 
We denote the set of all simulation traces of $\model$ by $\mathcal{L}(\model)$.

\subsection{Signal Temporal Logic}
\label{sec:mtl:intro}

Signal Temporal Logic (STL) was introduced as an extension to Metric Temporal Logic (MTL) \cite{BartocciEtAl2018survey} to reason about real-time properties of signals (simulation traces).
STL formulae are built over predicates on the variables of a signal using combinations of Boolean and temporal operators.  
The temporal operators include {\it eventually} $(\Diamond_\Ic)$, {\it always} $(\Box_\Ic)$ and {\it until} $(\Uc_\Ic)$,
where $\Ic$ encodes timing constraints.  

In this work, we interpret STL formulas over the observable simulation traces of a given system.
STL specifications can describe the usual properties of interest in system design such as (bounded time) {\bf reachability}, e.g., {\it between time 1 and 5, $\x$ should drop below $-10$}: $\Diamond_{[1,5)}\ ( \x \leq -10)$, and {\bf safety}, e.g., {\it after time 2, $\x$ should always be greater than $10$}: $\Box_{[2,+\infty)} (\x \geq 10)$.  
Beyond the usual properties, STL can capture sequences of events, e.g., $\Diamond_{\Ic_1} (\pi_1 \wedge \Diamond_{\Ic_2} (\pi_2 \wedge \Diamond_{\Ic_3} \pi_3))$, and infinite behaviors, e.g., {\bf periodic} behaviors : $\Box (\pi_1 \rightarrow \Diamond_{[0,2]} \pi_2)$, where $\pi_i$ are predicates over signal variables.

Informally speaking, we allow predicate expressions to capture arbitrary constraints over the state variables, inputs and parameters of the system.
In other words, we assume that predicates $\pi$ are expressions built using the grammar $\pi ::= f(\x,\u,\p)\geq c \; | \; \neg \pi_1 \; | \; (\pi) \; | \; \pi_1 \vee \pi_2 \; | \; \pi_1 \wedge \pi_2$, where $f$ is a function and $c$ is a constant in $\reals$.
Effectively, each predicate $\pi$ represents a subset in the space $\X \times \U \times \P$.
In the following, we represent that set that corresponds to the predicate $\pi$ using the notation $\Oc(\pi)$. 
For example, if $\pi = (x^{(1)} \leq -10) \vee (x^{(1)} +  x^{(2)} \geq 10)$ and we represent by $x^{(i)}$ the $i$-th component of the vector $\x$, then $\Oc(\pi ) = (\infty,-10] \times \reals \cup \{x \in \reals^2 \; | \; x^{(1)} +  x^{(2)} \geq 10 \}$. 

\begin{definition}[STL Syntax] 
	Assume $\Pi$ is the set of predicates and $\Ic$ is any non-empty interval of $\copreals$.
	The set of all well-formed STL formulas is inductively defined as
	\[ \varphi \; ::= \; \top \; | \; \pi \; | \; \neg \phi \; | \; \phi_1 \vee \phi_2 \; | \; \bigcirc\phi\; | \;\phi_1 U_\Ic \phi_2,\]
	where $\pi$ is a predicate, $\top$ is \emph {true}, $\bigcirc$ is Next and $U_\Ic$ is Until operator.
\end{definition}

For STL formulas $\psi$, $\phi$, we define 
\begin{alignat*}
\psi\wedge\phi&\equiv\neg(\neg\psi\vee\neg\phi),&&\\
\bot&\equiv\neg\top\ &&\text{(False)},\\
\psi\rightarrow\phi&\equiv\neg\psi\vee\phi\ &&\text{(}\psi\ \text{Implies}\ \phi\text{)},\\
\Diamond_I\psi&\equiv\top U_I\psi\ &&\text{(Eventually }\psi\text{)},\\
\Box_I\psi&\equiv\neg\Diamond_I\neg\psi\ &&\text{(Always }\psi\text{)},\\
\psi R_I\phi&\equiv\neg(\neg\psi U_I\neg\phi)\ &&\text{(}\psi\text{ Releases }\phi\text{)}
\end{alignat*}
using syntactic manipulation.

In our previous work \cite{FainekosP06fates}, we proposed robust semantics for STL formulas.  
Robust semantics (or robustness metrics) provide a real-valued measure of satisfaction of a formula by a trace in contrast to the Boolean semantics that just provide a {\it true} or {\it false} valuation.  
In more detail, given a trace $T$ of the system, its robustness w.r.t. a temporal property $\varphi$, denoted $\robsemD{\varphi}(T)$ yields a positive value if $T$ satisfies $\varphi$ and a negative value otherwise.
Moreover, if the trace $T$ satisfies the specification $\phi$, then the robust semantics evaluate to the radius of a neighborhood such that any other trace that remains within that neighborhood also satisfies the same specification.
The same holds for traces that do not satisfy~$\phi$.

\begin{definition} [STL Robust Semantics] 
	Given a metric $\genMet$, trace $\tss$ and $\Oc : \Pi \rightarrow 2^{\X \times \U \times \P}$, the robust semantics of any formula $\phi $ w.r.t $\tss$ at time instance $i\in N$ is defined as:
	\begin{align*} 
	\allowdisplaybreaks
	\dle \top\dri_{\genMet} (\tss,i) := &  +\infty  
	\displaybreak[2] \\
	\dle \pi \dri_{\genMet} (\tss,i) := & \left\{ \begin{array}{ll}
	- \inf\{\genMet((\x_i,\u_i,\p_i),\y)\;|\; \y \in \Oc(\pi) \} & \mbox{ if } (\x_i,\u_i,\p_i) \not \in \Oc(\pi)\\
	\inf\{\genMet((\x_i,\u_i,\p_i),\y)\;|\; \y \in \overline{\Oc(\pi)} \} & \mbox{ if } (\x_i,\u_i,\p_i) \in \Oc(\pi)\\
	\end{array} \right. 
	\displaybreak[2] \\
	\dle \neg \phi \dri_{\genMet}  (\tss,i)  :=  & - \dle \phi \dri_{\genMet}  (\tss,i) 
	\displaybreak[2] \\
	\dle \phi_1 \vee \phi_2 \dri_{\genMet}  (\tss,i) := &  \max\big(\dle \phi_1 \dri_{\genMet}  (\tss,i) , \dle \phi_2 \dri_{\genMet}  (\tss,i) \displaybreak[2]\big)  \\
	\dle \bigcirc \phi \dri_{\genMet}  (\tss,i)  :=  & 
	\displaybreak[2] \left\{ \begin{array}{ll}
	\dle \phi \dri_{\genMet}  (\tss,i+1) & \mbox{ if } i+1\in N\\
	-\infty  & \mbox{ otherwise }  \\
	\end{array} \right.\\
	\dle \phi_1 U_\Ic \phi_2 \dri_{\genMet}  (\tss,i) := & \max_{j \mbox{ s.t. } (t_j-t_i) \in \Ic} \bigg( \min\Big(\dle \phi_2 \dri_{\genMet}  (\tss,j) , \min_{i \leq k <j} \dle \phi_1 \dri_{\genMet}  (\tss,k) \Big)\bigg)
	\end{align*}
	\label{def:mitlrob}
\end{definition}

A trace $\tss$ satisfies an STL formula $\phi$ (denoted by $\tss \models\phi$), if $\dle \phi\dri_{\genMet} (\tss,0)>0$.
On the other hand, a trace $\tss'$ falsifies $\phi$ (denoted by $\tss' \not\models\phi$), if $\dle \phi\dri_{\genMet} (\tss',0)<0$.
Algorithms to compute $\robsemD{\varphi}$ have been presented in \cite{FainekosP06fates,FainekosSUY12acc,DonzeM10formats}.

\begin{figure}[htb]
	\begin{center}
		\includegraphics[width=0.7\linewidth]{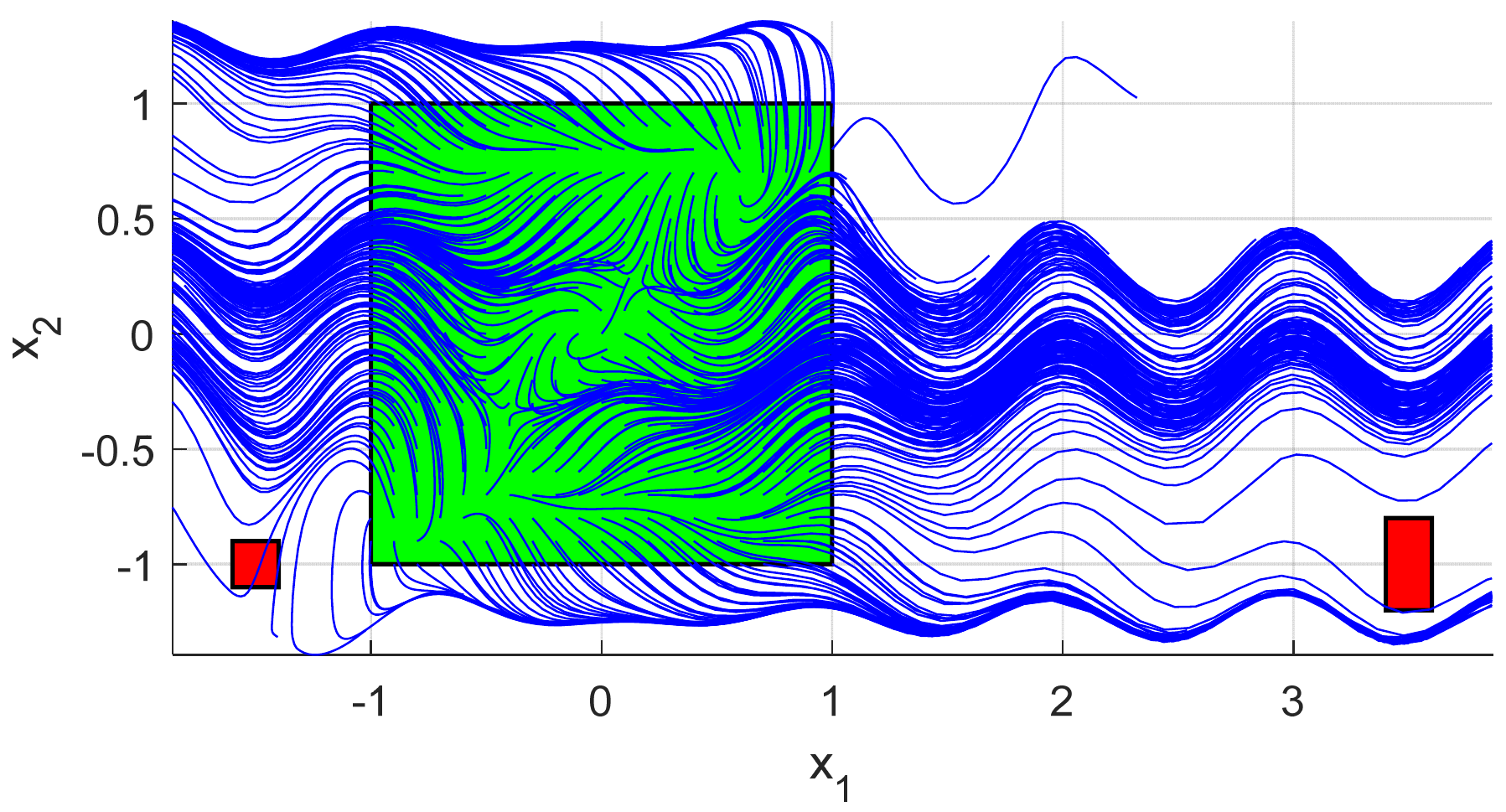} 
	\end{center}
	\caption{System trajectories.}
	\label{fig:simple:traj}
\end{figure}

\begin{figure}[htb]
	\begin{center}
		\includegraphics[width=0.7\linewidth]{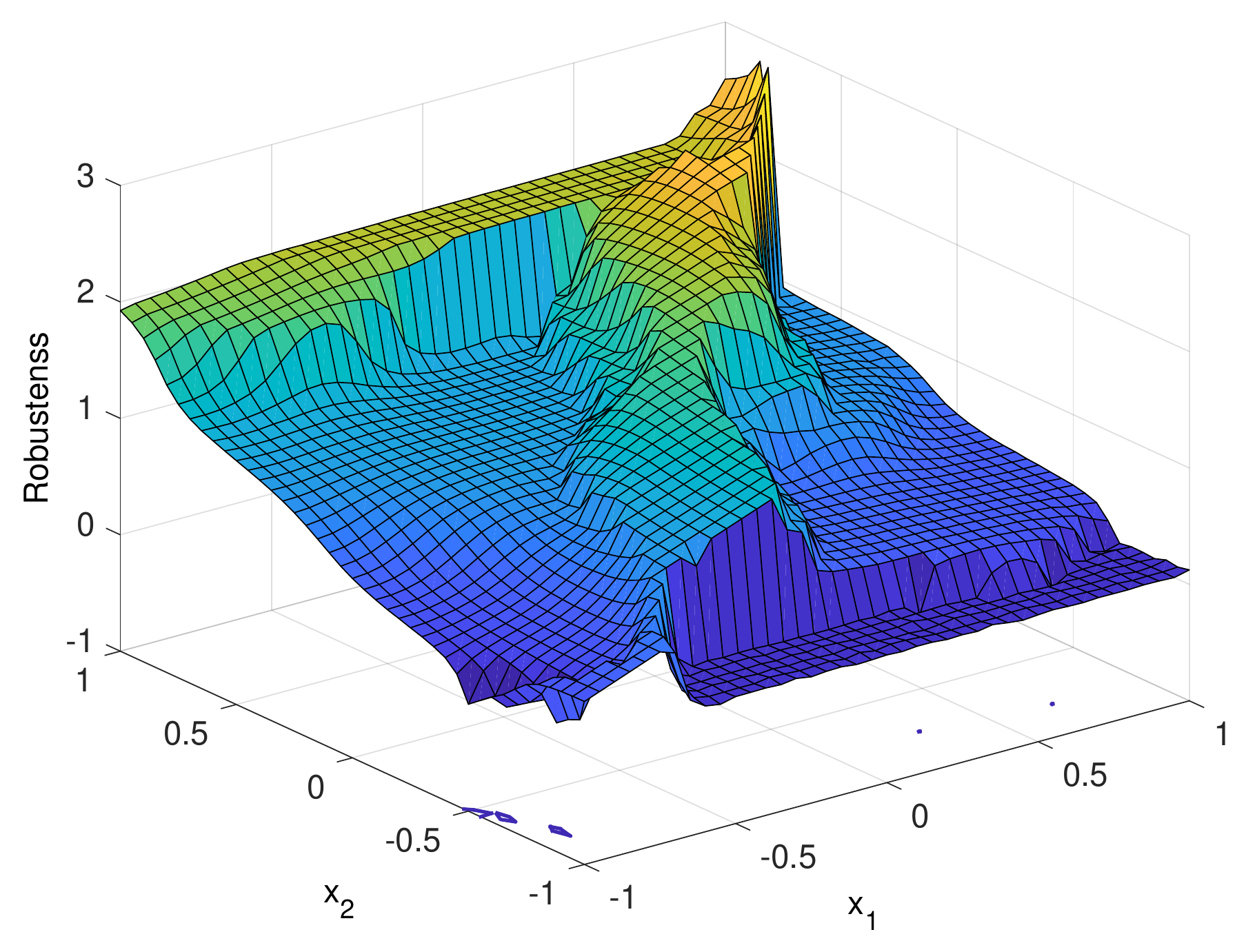} 
	\end{center}
	\caption{The resulting robustness landscape for specification~$\varphi$.}
	\label{fig:simple:rob}
\end{figure}

\begin{exmp}
	\label{exmp:simple}
	Consider the following dynamical system:
	\begin{align*}
	\dot x_1 & = x_1 - x_2 + 0.1t\\
	\dot x_2 & = x_2 \cos(2\pi x_2) - x_1 \sin(2\pi x_1) + 0.1 t
	\end{align*}
	Sample system trajectories with initial conditions over a grid of 0.05 intervals in each dimension over the set of initial conditions $[-1,1]^2$ are presented in Fig. \ref{fig:simple:traj}.
	Using the  specification 
	\begin{align*}
		\varphi = &\Box \neg (x \in [-1.6,-1.4]\times[-1.1,-0.9]) \\
		&\wedge \Box \neg (x \in [3.4,3.6]\times[-1.2,-0.8]),
	\end{align*}
	we can construct the robustness surface in Fig. \ref{fig:simple:rob}.
	Informally, the specification states that all system trajectories must always not enter any of the red boxes in Fig. \ref{fig:simple:traj}.
	Therefore, any falsifying system behavior will enter either of the red boxes.
	Note that the regions of the initial conditions that initiate falsifying trajectories correspond to negative values in the robustness landscape in Fig. \ref{fig:simple:rob}.
	
\end{exmp}

\subsection{Robustness-Guided Model Checking (RGMC)}

The goal of a \emph{model checking} algorithm is to ensure that all traces satisfy the requirement.
The robustness metric can be viewed as a fitness function that indicates the degree to which individual executions of the system satisfy the requirement~$\varphi$, with positive values indicating that the execution satisfies $\varphi$. 
Therefore, for a given system $\model$ and a given requirement~$\varphi$, the model checking problem is to ensure that for all $T\in \mathcal{L}(\model)$, $\robsemD{\varphi}(T)>0$.

Let $\varphi$ be a given STL property that the system is expected to satisfy.  
The robustness metric $\robsemD{\varphi}$ maps each simulation trace $T$ to a real number $r$.  
Ideally, for the STL verification problem, we would like to prove that $\inf_{y \in \mathcal{L}(\Sigma)} \mathcal{R}_{\varphi}(y) > \varepsilon > 0$ where $\varepsilon$ is a desired robustness threshold.  

\vspace*{0.5cm}

\vspace*{0.5cm}

\subsection{Falsification and Critical System Behaviors} \label{sec:critical}

In this work, we focus on identifying critical counterexamples, which we refer to as \emph{glancing} examples.
For the autonomous driving system, and its corresponding requirement, described below, robustness values that are negative and large in magnitude will correspond to traces where the ego vehicle collides with an object at high velocity.
Glancing counterexamples are those that correspond to robustness values that are close to $0$, where the ego vehicle collides with an object at very low speeds.
Glancing counterexamples are valuable for designers, as they provide examples of behaviors at the boundary between satisfying and falsifying the given property \cite{TuncaliPF16itsc}.
As discussed in \cite{shalev2017formal}, it is important to extract the cases where the collision is avoidable, instead of failures caused by a pedestrian or an agent vehicle making a maneuver leading to an unavoidable collision.

To identify critical system behaviors, we leverage existing work on \emph{falsification}, which is the process of identifying system traces $T$ that do not satisfy $\varphi$.
For the STL falsification problem, falsification attempts to solve the problem:
Find $T\in \mathcal{L}(\Sigma)$ s.t. $\robsemD{\varphi}(T) < 0$. 
This is done using best effort solutions to the following optimization problem: 
$T^\star = \arg \min_{T \in \mathcal{L}(\Sigma)}  \robsemD{\varphi}(T).$
If $ \robsemD{\varphi}(T^\star)<0$, then a counterexample (adversarial sample) has been identified, which can be used for debugging or for training.
In order to solve this non-linear non-convex optimization problem, a number of stochastic search optimization methods can be applied (e.g., \cite{AbbasFSIG13tecs} -- for an overview see \cite{HoxhaEtAl14difts,KapinskiEtAl2016csm}).

We use falsification methods to identify glancing examples for an autonomous driving system, where we identify 
\begin{equation}
\hat{T}^\star = \arg \min_{T \in \mathcal{L}(\Sigma)} | \robsemD{\varphi}(T) |
\label{eqn:T_star}
\end{equation}
as the glancing instance.
The minimum of $| \robsemD{\varphi}(T) |$ for the autonomous driving system, which is usually $0$, can be designed to correspond to a trace where a collision occurs with a small velocity.

\subsection{Covering Arrays}
\label{sec:covering_arrays}

In software systems, there can often be a large number of discrete input parameters that affect the execution path of a program and its outputs.
The possible combinations of input values can grow exponentially with the number of parameters.
Hence, exhaustive testing on the input space becomes impractical for fairly large systems.
A fault in such a system with $k$ parameters may be caused by a specific combination of $t$ parameters, where $1 \leq t \leq k$.
One best-effort approach to testing is to make sure that all combinations of any $t$-sized subset (i.e., all $t$-way combinations) of the inputs is tested.

A \textit{covering array} is a minimal number of test cases such that any $t$-way combination of test parameters exist in the list \cite{hartman2005software}.
Covering arrays are generated using optimization-based algorithms with the goal of minimizing the number of test cases.
We denote a $t$-way covering array on $k$ parameters by $CA(t, k, (v_1, ..., v_k))$, where $v_i$ is the number of possible values for the $i^{th}$ parameter.
The size of covering array increases with increasing $t$, and it becomes an exhaustive list of all combinations when $t=k$.
Here, $t$ is considered as the \textit{strength} of the covering array.
In practice, $t$ can be chosen such that the generated tests fit into the testing budget.
Empirical studies on real-world examples show that more than $90$ percent of the software failures can be found by testing 2 to 4-way combinations of inputs \cite{kuhn2013introduction}.

Despite the $t$-way combinatorial coverage guaranteed by covering arrays, a fault in the system possibly may arise as a result of a combination of a number parameters larger than $t$.
Hence, covering arrays are typically used to supplement additional testing techniques, 
like uniform random testing.
We consider that because of the nature of the training data or the network structure, NN-based object detection algorithms may be sensitive to a certain combination of properties of the objects in the scene.
In Sec. \ref{sec:applications}, we describe how \toolname combines covering arrays to explore discrete and discretized parameters with falsification on continuous parameters.

\section{Framework} \label{sec:framework}

We describe \toolname, a framework for performing testing and analysis of autonomous driving systems in a virtual environment.
The simulation environment used in \toolname includes a vehicle perception system, a vehicle controller, a model of the physical environment, and a mechanism to render 2D images from 3D models.
The framework uses freely available and low cost tools and can be run on a standard desktop PC.
Later, we demonstrate how \toolname can be used to implement traditional testing approaches, as well as advanced automated testing approaches that were not previously possible using existing frameworks.

\begin{figure}[htb]
\begin{centering}
\begin{tikzpicture}[auto, node distance=5cm,>=latex',scale=0.6, every node/.style={scale=0.6}]

\tikzset{
  font={\fontsize{16pt}{16}\selectfont}}

\tikzstyle{block} = [draw, font=\Large, fill=green!20, rectangle, 
    minimum height=4em, minimum width=6em,align=center,execute at begin node=\setlength{\baselineskip}{1.5em}]
\tikzstyle{texts} = [minimum height=4em, minimum 	width=6em,align=center,execute at begin node=\setlength{\baselineskip}{1.5em}]
\tikzstyle{input} = [coordinate]
\tikzstyle{output} = [coordinate]
    
    \node [draw, rectangle, rounded corners=5pt, minimum width=21.5cm, minimum height=6.25cm, fill=black!10,label={[anchor=north west, inner sep=10pt]north west:Simulation Environment}] (sim env) at (7.75,0.75) {};
    
	\node [draw, rectangle, rounded corners=5pt,xshift=0.25cm, minimum width=10.0cm, minimum height=4.5cm, fill=blue!10,label={[anchor=north west, inner sep=10pt]north west:Webots}] (webots) at (12.5,0.75) {};    
    
	\node [draw, rectangle, rounded corners=5pt, minimum width=5cm, minimum height=4.5cm, fill=blue!10,label={[anchor=north west, inner sep=10pt]north west:Tensorflow}] (tensorflow) at (0.5,0.5) {};

	\node [draw, rectangle, rounded corners=5pt, minimum width=4cm, minimum height=3cm, fill=red!10,label={[anchor=north west, inner sep=10pt]north west:SqueezeDet}] (squeezedet) at (0.55,0.25) {};

    \node [block,xshift=0.5cm,text width=8em](perception){\textsc{Perception System}} ;
    
    \node [block,right of=perception,text width=9em] (controller) {\textsc{Controller}};

  \node [block,right of=controller,xshift=0.25cm,text width=14em] (environment) {\textsc{Vehicle and \\Environment Model}};

\node [block,right of=environment, xshift=0.1cm,text width=8em] (renderer) {\textsc{Renderer}};

    
 \draw [very thick,->] (perception) --  node [name=r] {}(controller);
  \draw [very thick,->] (controller) -- node [name=m] {} (environment);
   \draw [very thick,->] (environment) -- node [name=out] {} (renderer);

\draw [very thick,->] ($(renderer.east)$) -- ($(renderer.east)+(0.6,0)$) -- ($(renderer.east)+(0.6,-2)$) -| (perception.south) ;

\end{tikzpicture}
\caption{Overview of the simulation environment. \label{fig:framework}}
\end{centering}
\end{figure}
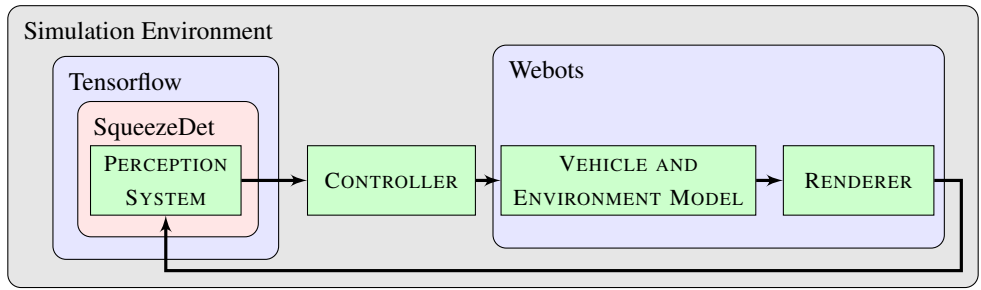

Fig. \ref{fig:framework} shows an overview of the simulation environment.
The simulation environment has three main components: the perception system, the controller, and the environment modeling framework, which also performs image rendering.

For the perception system, we feed the vehicle front camera images to a lightweight Deep NN, SqueezeDet, which performs object detection and classification~\cite{wu2016squeezedet}.
SqueezeDet is implemented in TensorFlow\texttrademark \cite{abadi2016tensorflow}, and it outputs a list of object detection boxes with corresponding class probabilities.
This network was originally trained on real image data from the KITTI dataset \cite{geiger2012we} to achieve accuracy comparable to the popular AlexNet classifier \cite{krizhevsky2012imagenet}.
We further train this network on the virtual images generated in our framework.

For our experiments, we use only camera information and neglect other sensors that would typically be available on autonomous automobile systems, such as LIDAR and radar.
In principle, this means that the controller that we use is likely more sensitive to the performance of the image-based perception systems, as compared to a typical autonomous driving system, since the computer vision system is essentially the only sensor information that we use.
This assumption was made for practical considerations, but we argue that this is acceptable for our prototype framework, as this allows us to focus our attention on the NN-based computer vision system, which is often considered to be the most difficult aspect of autonomous driving systems to test.
Having said that, we realize that this assumption is significant; 
future work will include adding models of other typical sensing systems, such as LIDAR and radar.

For our evaluations of the proposed testing framework, we implemented a simple  \emph{collision avoidance} controller in Python. 
The controller takes the detection box and object class information from SqueezeDet and estimates the actual positions of the detected objects (e.g., pedestrians or vehicles) by a linear mapping of the detection box pixel positions and sizes to the 3D space.
The controller also utilizes the Median Flow tracker  \cite{kalal2010forward} implementation in OpenCV to track and estimate future positions of the objects.
A collision is predicted if an object is, or in the future will be, in front of the controlled vehicle at a distance shorter than a threshold.
When there is no collision risk, the controller drives the car with a constant throttle.
When a future collision is predicted, it applies the brakes at the maximum level, and if the predicted time to collision is less than a threshold, it also steers away from the object to avoid a possible collision.
We note that \toolname does not restrict the controller; our controller can be easily exchanged with an alternative.

The environment modeling framework is implemented in Webots \cite{michel2004cyberbotics}, a robotic simulation framework that models the physical behavior of robotic components, such as manipulators and wheeled robots, and can be configured to model autonomous driving scenarios.
In addition to modeling the physics, a graphics engine is used to produce images of the scenarios.
In \toolname, the images rendered by Webots are configured to correspond to the image data captured from a virtual camera that is attached to the front of a vehicle.

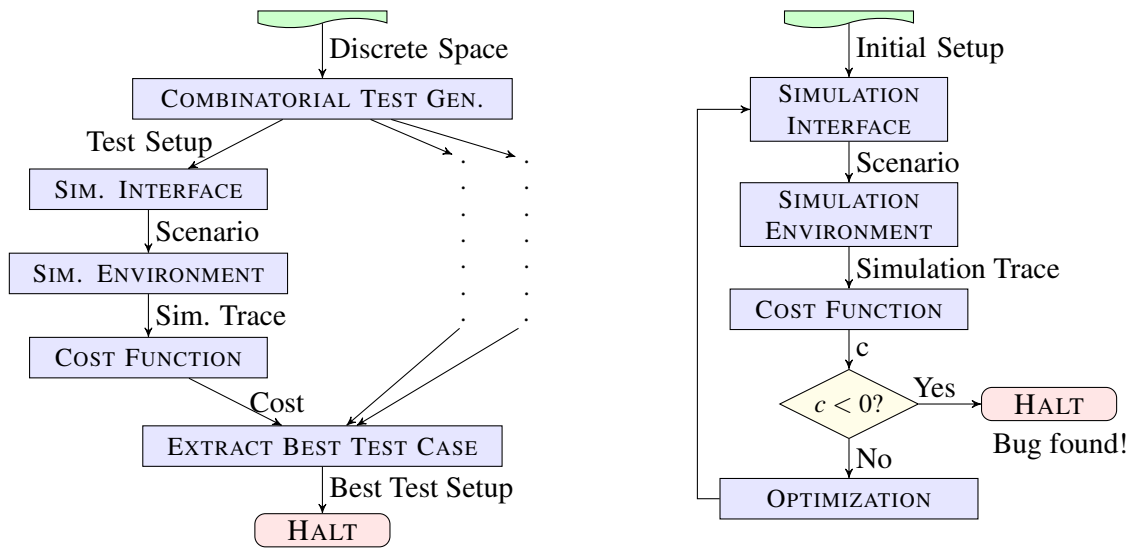
\begin{figure}[htb]
	\begin{centering}
		\begin{tikzpicture}[>=stealth',scale=0.7, every node/.style={scale=0.7}]

\tikzset{
  font={\fontsize{16pt}{16}\selectfont}}

\tikzstyle{stext}=[font=\fontsize{14}{14}\selectfont]
\tikzstyle{smalltext}=[font=\fontsize{14}{14}\selectfont]
\tikzstyle{block}=[draw,fill=white,rectangle,minimum height=2em,text
                   width=4.5em,fill=blue!10,stext,inner sep=4pt,align=center,execute at begin node=\setlength{\baselineskip}{1.2em}]
\tikzstyle{file}=[tape,fill=red!10,tape bend top=none,draw,inner
                  sep=2pt,text width=5em,align=center,smalltext,execute at begin node=\setlength{\baselineskip}{1.2em}]
\tikzstyle{note} = [coordinate]

\node[file,fill=green!20, text width=6em]  (init setup) {};

\node[block,below of= init setup, node distance=18mm, text width=9em] (sim int) {\textsc{Simulation \\Interface}};

\node[block,below of=sim int, node distance=20mm, text width=10em] (s) {\textsc{Simulation \\Environment}};

\node[block,below of=s,node distance=18mm,text width=11em] (cost) {\textsc{Cost Function}};

\node[draw,smalltext,diamond,aspect=2,text width=5em,below of=cost,fill=yellow!10,node distance=18mm,inner sep=0pt,align=center] (decision) 
     {$c<0$?};

\node[block,below of=decision,node distance=18mm,text width=12em] (optim) {\textsc{Optimization}};

\node[draw,rounded corners,align=center, right of= decision, node distance=38mm,fill=red!10,text width=6em] (stop) 
     {\textsc{Halt}};
\node [note,name=bugfound, font=\Large, below right=0.6cm and -0.75cm of stop,label={Bug found!}] {};

\draw[->] (init setup) -- node[name=initalize, right, text width=9em,execute at begin node=\setlength{\baselineskip}{1.2em}] {Initial Setup} (sim int);

\draw[->] (sim int) -- node[name=initalize, right, text width=6em,execute at begin node=\setlength{\baselineskip}{1.2em}] {Scenario} (s);

\draw[->] (s) -- node[name=simresults, right]{Simulation Trace} (cost);

\draw[->] (cost) -- node[name=costout, right]{c} (decision);

\draw[->] (decision) -- node[name=no,right]{No} (optim);

\draw[->] (decision) -- node[name=yes,above, near start]{Yes} (stop);

\draw [->] (optim.west) -| node[name=newparam, left, near end, text width=3em,execute at begin node=\setlength{\baselineskip}{1.2em}]{} ($(sim int.west)+(-1.0,0.0)$) -- (sim int.west) ;

\node[file, left of=init setup, node distance = 100mm, fill=green!20, text width=6em]  (discrete init) {};
\node[block, below of=discrete init, node distance=16mm, text width=18em] (combinatorial) {\textsc{Combinatorial Test Gen.}};
\draw[->] (discrete init) -- node[name=discrete space, right, text width=12em,execute at begin node=\setlength{\baselineskip}{1.2em}] {Discrete Space} (combinatorial);

\node[block, below left=1.7em and -18mm of combinatorial, text width=11em] (sim int discrete) {\textsc{Sim. Interface}};
\draw[->] (combinatorial) -- node[name=setup discrete, left, text width=7em,execute at begin node=\setlength{\baselineskip}{1.2em}] {Test Setup} (sim int discrete);

\node[stext, below right=4.5mm and -8mm of combinatorial, text width=1em] (s1) {\textsc{.}};
\draw[->] (combinatorial) -- node[name=setup discrete 2, left, text width=7em,execute at begin node=\setlength{\baselineskip}{1.2em}] { } (s1);
\node[stext, below of=s1, node distance=5mm, text width=1em] (s3) {\textsc{.}};
\node[stext, below of=s3, node distance=5mm, text width=1em] (s4) {\textsc{.}};
\node[stext, below of=s4, node distance=5mm, text width=1em] (s5) {\textsc{.}};
\node[stext, below of=s5, node distance=5mm, text width=1em] (s6) {\textsc{.}};
\node[stext, below of=s6, node distance=5mm, text width=1em] (s7) {\textsc{.}};
\node[stext, below of=s7, node distance=5mm, text width=1em] (s11) {\textsc{.}};

\node[stext, right of=s1, node distance=12mm, text width=1em] (s2) {\textsc{.}};
\draw[->] (combinatorial) -- node[name=setup discrete 2, left, text width=7em,execute at begin node=\setlength{\baselineskip}{1.2em}] { } (s2);
\node[stext, below of=s2, node distance=5mm, text width=1em] (s3) {\textsc{.}};
\node[stext, below of=s3, node distance=5mm, text width=1em] (s4) {\textsc{.}};
\node[stext, below of=s4, node distance=5mm, text width=1em] (s5) {\textsc{.}};
\node[stext, below of=s5, node distance=5mm, text width=1em] (s6) {\textsc{.}};
\node[stext, below of=s6, node distance=5mm, text width=1em] (s7) {\textsc{.}};
\node[stext, below of=s7, node distance=5mm, text width=1em] (s12) {\textsc{.}};

\node[block, below of=sim int discrete, node distance=16mm, text width=13em] (sim env discrete) {\textsc{Sim. Environment}};
\draw[->] (sim int discrete) -- node[name=setup discrete, right, text width=16em,execute at begin node=\setlength{\baselineskip}{1.2em}] {Scenario} (sim env discrete);

\node[block, below of=sim env discrete, node distance=16mm, text width=11em] (cost discrete) {\textsc{Cost Function}};
\draw[->] (sim env discrete) -- node[name=simresults discrete, right]{Sim. Trace} (cost discrete);

\node[block, below of=combinatorial, node distance=66mm, text width=17em] (minimum discrete) {\textsc{Extract Best Test Case}};
\draw[->] (cost discrete) -- node[name=simresults discrete, right]{\ Cost} (minimum discrete);
\draw[->] (s11) -- node[name=simresults discrete, right]{} (minimum discrete);
\draw[->] (s12) -- node[name=simresults discrete, right]{} (minimum discrete);

\node[draw, rounded corners ,align=center, below of=minimum discrete, node distance=16mm,fill=red!10,text width=6em] (stop discrete) 
{\textsc{Halt}};
\draw[->] (minimum discrete) -- node[name=best result discrete, right]{Best Test Setup} (stop discrete);

\end{tikzpicture}
		\caption{Flowcharts illustrating the combinatorial testing (left) and falsification (right) approaches. \label{fig:flowchart}}
	\end{centering}
\end{figure}

The process used by \toolname for test generation and execution for discrete and discretized continuous parameters is illustrated by the flowchart shown in Fig. \ref{fig:flowchart} (left).
\toolname first generates test cases that correspond to scenarios defined in the simulation environment using covering arrays as a combinatorial test generation approach.
The scenario setup is communicated to the simulation interface using TCP/IP sockets.
After a simulation is executed, the corresponding simulation trace is received via socket communication and evaluated using a cost function.
Among all discrete test cases, the most promising one is used as the initial test case for the falsification process shown in Fig. \ref{fig:flowchart} (right).
For falsification, the result obtained from the cost function is used in an optimization setting to generate the next scenario to be simulated.
For this purpose, we used S-TaLiRo \cite{FainekosSUY12acc}, which is a \matlab toolbox for falsification of CPSs.
Similar tools, such as Breach \cite{DonzeM10formats}, can also be used in our framework for the same purpose.

\section{Testing Application} \label{sec:applications}
In this section, we present the results from a case study using an autonomous driving scenario to evaluate \toolname.
We describe a specific driving scenario and provide the corresponding STL specification.
Then, we describe the testing procedure that we use to identify critical (glancing) system behaviors, which is based on covering arrays and robustness-guided falsification. Finally, we analyze the results.

\subsection{Scenario Setup} \label{sec:scenario}
In this section, we describe the driving scenario that we use to evaluate \toolname.
The scenario is selected to be both challenging for the autonomous driving system and also analogous to plausible driving scenarios experienced in real world situations.
In general, the system designers will need to identify crucial driving scenarios, based on intuition about challenging situations, from the perspective of the autonomous vehicle control system. A thorough simulation-based testing approach will include a wide array of scenarios that exemplify critical driving situations.

\begin{figure}[htb]
	\begin{centering}
		\includegraphics[width=0.7\columnwidth]{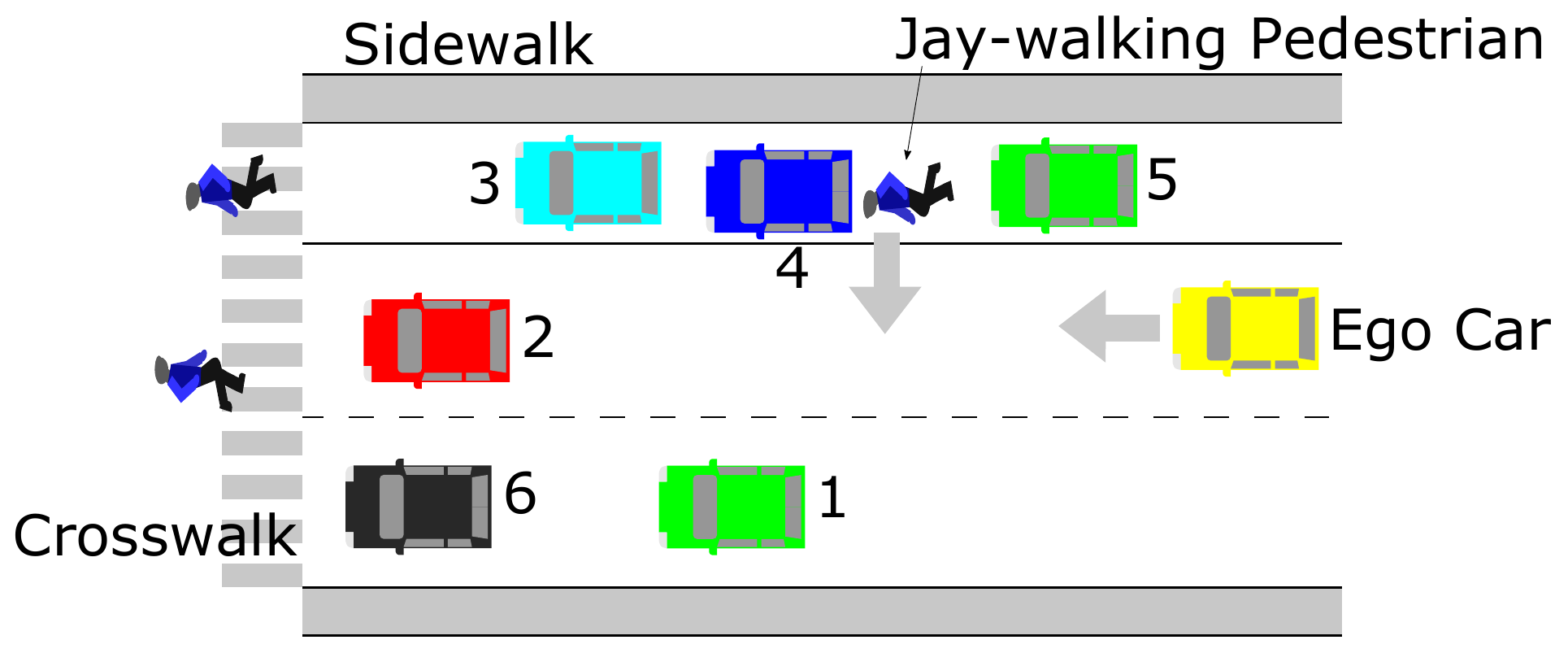}
		\caption{Overview of the scenario used for evaluations.}
		\label{fig:scenariooverview}
	\end{centering}
\end{figure}

We use the scenario depicted in Fig. \ref{fig:scenariooverview}, which we call system $\modeld$.
The scenario includes the ego car on a one-way road, with parked cars in lanes on the left and right of the vehicle.
In the Figure, agent Vehicles 1, 3, 4, 5, and 6 are parked.
Also, Vehicle 2 is stopped due to the two pedestrians in the crosswalk.
The Jay-walking Pedestrian is crossing the street between Vehicles 4 and 5. This pedestrian crosses in front of the Ego car until it reaches the leftmost lane, at which time it changes direction and returns to the right Sidewalk.

Several aspects of the scenario depicted in Fig. \ref{fig:scenariooverview} are parameterized, meaning that their values are fixed for any given simulation by appropriately selecting the model parameters $\p_d$.
The longitudinal position of Vehicle 1 is parameterized.
Also the colors of Vehicles 1 through 5 and the shirt and pants of the Jay-walking Pedestrian are parameterized and can each take one of 5 values: red, green, blue, white, or black. 
Additionally, Vehicles 1 through 5 can be one of 5 different car models.
Further, the walking speed of the Jay-walking Pedestrian is parameterized, along with the initial longitudinal position of the Ego car. Fog may be present in the scene, and this corresponds to a Boolean variable that determines whether the fog is present or not. In all, $\modeld$ has 13 discrete ($V=13$) and 3 continuous parameters ($W=3$).

\begin{figure}[htb]
	\centering
	\subfloat[Overview of testing scenario]{{\includegraphics[trim={0 0.23in 0 0},clip,width=0.7\columnwidth]{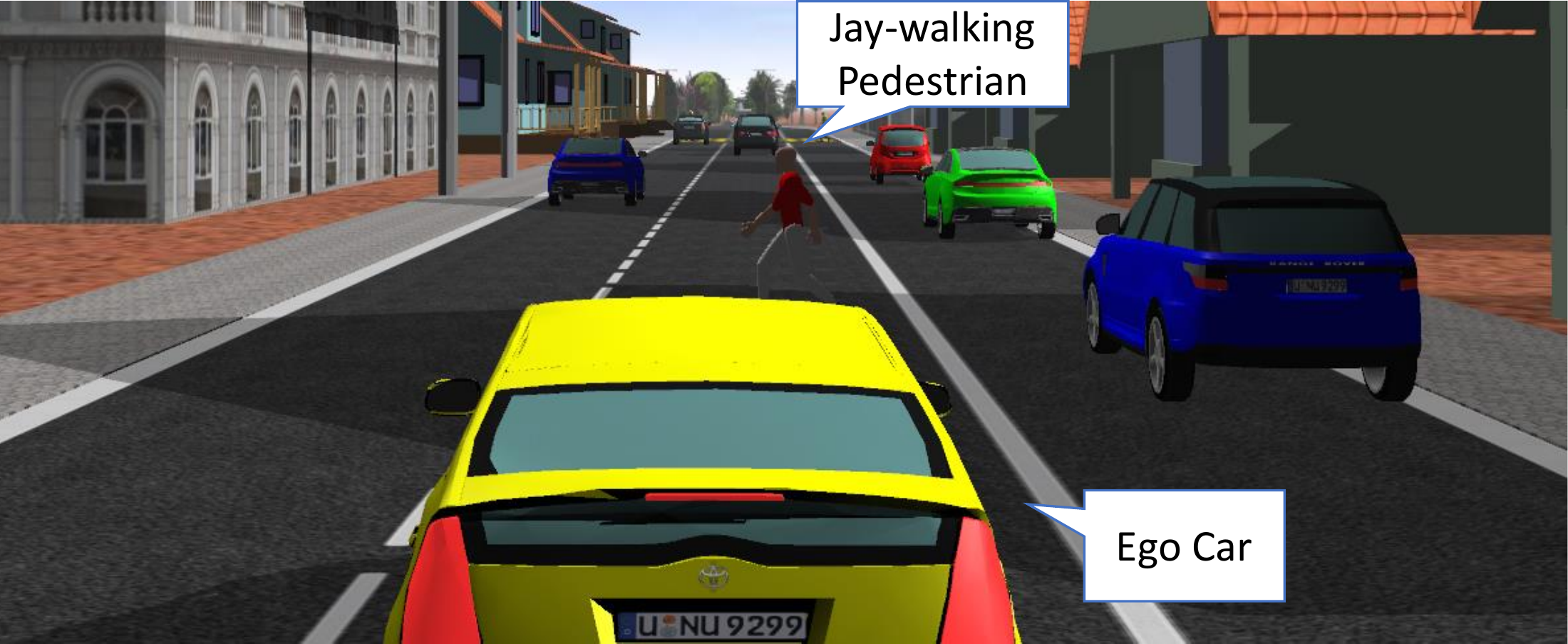}}\label{fig:imageexamples_sceneview}}
	\hfill
	\subfloat[Camera view and object detection and classification on the image]{{\includegraphics[width=0.7\columnwidth]{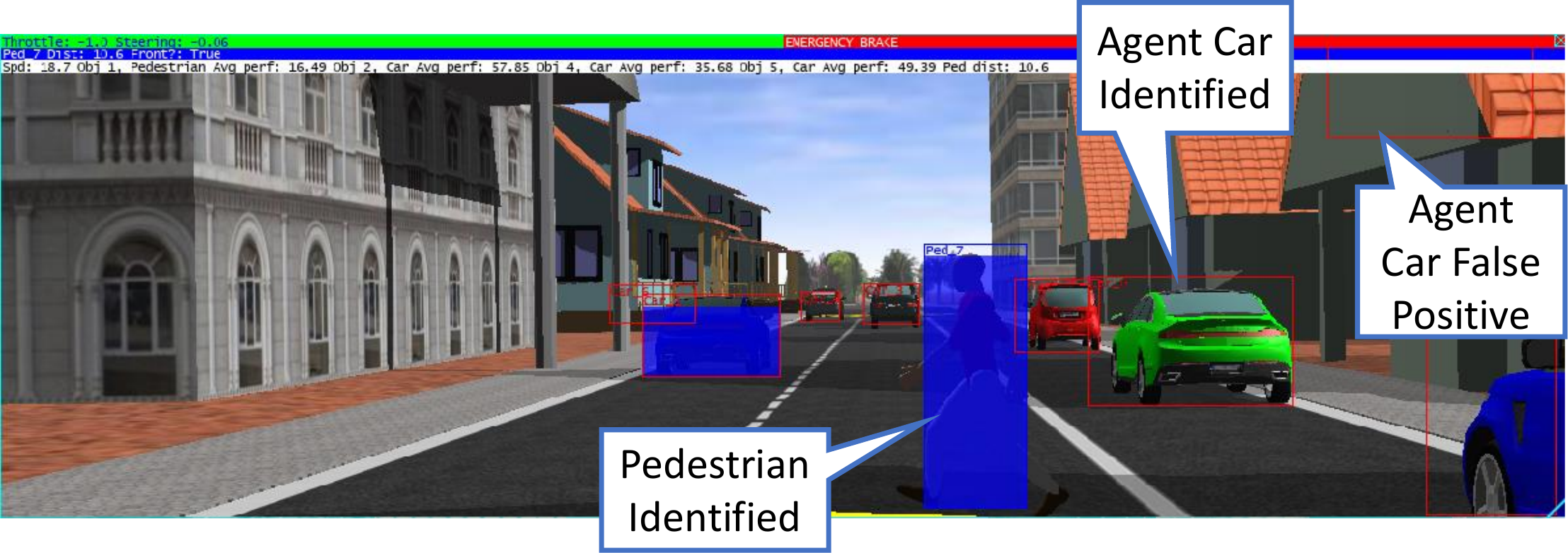}}\label{fig:imageexamples_cameraview}}
	\caption{Webots images from testing scenario.}
	\label{fig:imageexamples}
\end{figure}

Figure \ref{fig:imageexamples} shows examples of the scenario described above.
The images are generated by the Webots framework, which implements the physics model and provides the image rendering of the scene.
Figure~\ref{fig:imageexamples_sceneview} shows an example of the scene during a simulation run performed using the \toolname framework.
The Ego car is in the bottom-center of the image, and the Jay-walking Pedestrian can be seen just in front of the Ego car.
The parked cars can be seen in front of the Ego car on the left and right sides.

Figure \ref{fig:imageexamples_cameraview} shows an example of a camera image that is processed by the DNN, SqueezeDet. This image corresponds to the scene shown in Fig. \ref{fig:imageexamples_sceneview}. The object detection and classification performed by SqueezeDet is used to estimate where vehicles and pedestrians appear in the image, and the corresponding estimates are shown with red bounding boxes.  
The solid blue boxes indicate vehicles or pedestrians detected that the perception system determines are in danger of being collided with, unless action is taken by the controller.

\vspace*{1.0cm}

\subsection{STL Specifications} \label{sec:stl_specs_for_experiments}
Below is the STL specification used in our experiments:
\vspace*{-0.09in}

\[\phid = \Box (\pi_{v_{ego}, mov.}\hspace{-0.04in} \implies\hspace{-0.04in} (\neg \pi_{0, coll} \wedge ... \wedge \neg\pi_{l, coll} \wedge \neg\pi_{0, \overline{coll}} \wedge ... \wedge \neg\pi_{l, \overline{coll}}))\]
where, for all $i \in \{1, .., l\}$,
\begin{align*}
\pi_{v_{ego}, mov.} = & v_{ego} > \epsilon_{speed}\\
\pi_{i, coll} = & dist(i, ego) < \epsilon_{dist} \wedge \textit{front}(i, ego) = \top\\
\pi_{i, \overline{coll}} = & dist(i, ego) < 0.
\end{align*}

In the above specification, the Ego vehicle is indexed with $ego$, and all other objects in the environment are indexed from $1$ to $l$.
Position of an object is represented with $\mathbf{x}$.
$\textit{front}(i, ego)$ evaluates to true~($\top$) if object $i$ is partially in the front corridor of the Ego vehicle and to false ($\bot$) otherwise.
$\textit{dist}(i, ego)$ gives the minimum Euclidean distance between front bumper of the Ego vehicle and the object $i$.
The specification basically describes that the Ego vehicle should not hit another vehicle or pedestrian.
The predicates $\pi_{i, \overline{coll}}, \forall i \in \{1, ..., l\}$ are added so that the robustness corresponds to the  Euclidean distance of the corresponding object when it is not in front corridor of the Ego vehicle.

\subsection{Testing Strategies} \label{sec:strategies}
In this section, we describe three testing strategies that we implement in \toolname.
We compare their performance for the scenario described in Sec. \ref{sec:scenario}.
Because of the simple logic of our controller and the nature of the test scenario, it is relatively easy to find test cases that result in a bad behavior with respect to the system requirement;
however, finding test cases that correspond to behaviors that are on the boundary between safe and unsafe, that is, glancing cases, is a challenging task.
In our tests, we search for glancing cases (\textit{i.e.,} $\hat{T}^\star$, as defined in Eq. \ref{eqn:T_star} in Sec. \ref{sec:critical}), where a collision occurs at very low speeds or a collision is nearly missed.

\subsubsection{Global Uniform Random Search (\GUR)}
One naive approach to testing the autonomous driving system $\modeld$ against property $\phid$ is to select a collection of $J$ test cases (parameters $\p_{d_j}\in \P_d$, where $1 \leq j \leq J$) at random, run simulations to obtain the corresponding traces $T_{d_j}\in \mathcal{L}(\modeld)$, and then check whether the traces satisfy the specification, by checking whether $\robsemD{\phid}(T_{d_j}) > 0$ for all $1 \leq j \leq J$. 
To indentify the critical (glancing) cases using \GUR, we use $\arg \min_{1 \leq j \leq J} | \robsemD{\phid}(T_{d_j}) |$.
\GUR may not be efficient, as it is not guided towards critical areas.
The alternative methods below address this issue.

\subsubsection{Covering Arrays and Uniform Random Search (\CAUR)}
To provide guidance towards behaviors that demonstrate critical behaviors for system $\modeld$, we use a covering array, which is generated by the ACTS tool \cite{kuhn2013introduction}, for the discrete variables $\Pd_d$, and we combine this with a falsification approach that we use to explore the continuous variables $\Pc_d$.

The \CAUR method first creates a list of test cases $T_{d_j}$, where $1 \leq j \leq J$, and evaluates against the specification $\phid$.
The covering array is generated on the discrete variables $\Pd_d$ and discretized version of the continuous variables $\hat{\Pc}_d$. 

To identify a glancing behavior, the discrete parameters $\pd_{d_j}$ from the covering array that correspond to the $T_{d_j}$ that minimizes $| \robsemD{\phid}(T_{d_j}) |$ over $1 \leq j \leq J$ are used to create the test cases using the uniform random method over the continuous variables. 
$T_{d_k}$ that corresponds to the minimum value of $| \robsemD{\phid}(T_{d_j}) |$ is then taken as a glancing behavior.

The \CAUR method provides some benefits over the \GUR method, as it first uses covering arrays to identify promising valuations for the discrete variables. 
One problem with \CAUR is that the search over the continuous variables is still performed in an unguided fashion.
Next, we describe another approach that addresses this problem.

\subsubsection{Covering Arrays and Simulated Annealing (\CASA)}
The third method that we evaluate combines covering arrays to evaluate the discrete variables $\Pd_d$ and uses simulated annealing to search over the continuous variables $\Pc_d$.

The \CASA method is identical to the \CAUR method, except that, instead of generating continuous parameters $\pc_{d_k}$ offline, a cost function is used to guide a search of the continuous variables. 
To identify glancing behaviors, we use $| \robsemD{\phid}(T_{d_j}) |$ as the cost function.

\begin{figure}[htb]
	\begin{center}
		\begin{tabular}{c}
			\includegraphics[width=.23\columnwidth]{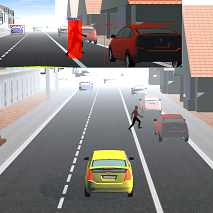}\hspace*{0in}
			\includegraphics[width=.23\columnwidth]{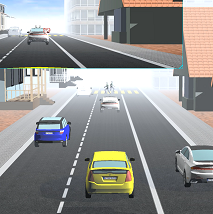}\hspace*{0in}
			\includegraphics[width=.23\columnwidth]{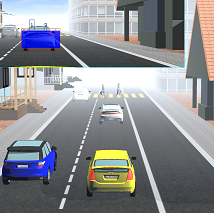}\hspace*{0in}
			\includegraphics[width=.23\columnwidth]{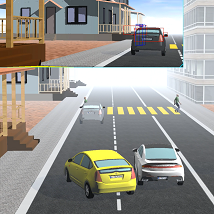}\hspace*{0in}			
			\vspace*{0.05cm}\\
			(a)\vspace*{0.25cm}\\
			\includegraphics[width=.23\columnwidth]{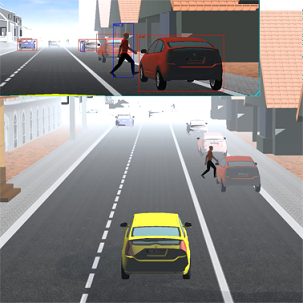}\hspace*{0in}
			\includegraphics[width=.23\columnwidth]{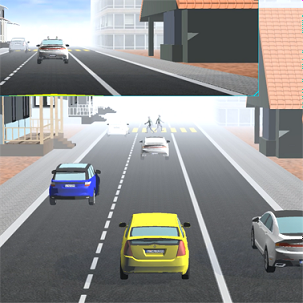}\hspace*{0in}
			\includegraphics[width=.23\columnwidth]{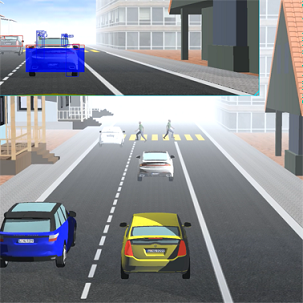}\hspace*{0in}
			\includegraphics[width=.23\columnwidth]{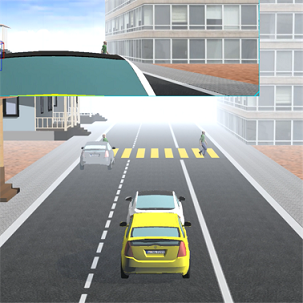}\hspace*{0in}
			\vspace*{0.05cm}\\
			(b)\\
		\end{tabular}
		\caption{Time-ordered images from (a) a non-failing test from the covering array and (b) a failure detected by falsification.}
		\label{fig:before_after_falsification}
	\end{center}
\end{figure}

Fig. \ref{fig:before_after_falsification}a illustrates a test run from a non-failing but close to failing covering array test.
In this example, even though there had been detection failures in the perception system in detecting the white vehicle ahead, they did not result in a collision, as the Ego vehicle was able to take a steering action and avoid the collision.
By utilizing Simulated Annealing over the continuous parameters, while keeping the discrete parameters unchanged, we can search for a collision and find a behavior like the one illustrated in Fig. \ref{fig:before_after_falsification}b, where the detection failures in the perception for the white vehicle ahead leads to a rear-end collision.

\vspace*{0.5cm}

\subsection{Experiment Results}

\begin{figure}[htb]
	\centering
	\includegraphics[width=0.6\linewidth]{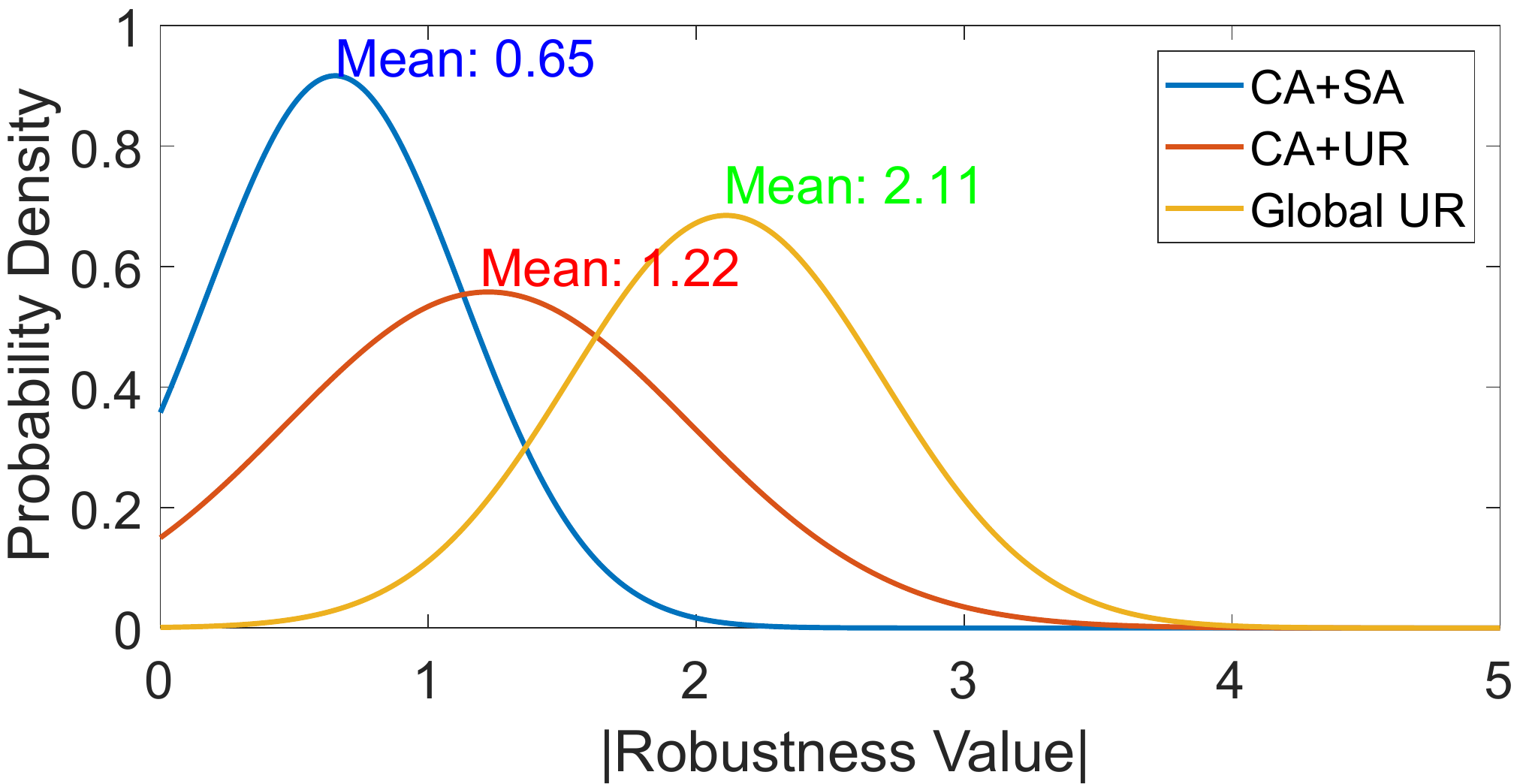}
	\caption{Probability distribution of the robustness values.}
	\label{fig:probability_distribution}
\end{figure}

We applied the testing strategies described in Sec. \ref{sec:strategies} to $\modeld$.
Fig. \ref{fig:probability_distribution} illustrates how the glancing-case robustness values fit to a truncated normal distribution for different test approaches, together with the mean robustness values.
The robustness values closer to zero represent a better performance for the related test approach, as they correspond to behaviors where a collision either nearly occurred or nearly did not occur.
For each test approach we perform 20 trials, each with a total budget of 200 simulations.
 
For \GUR, the average robustness values reported are the mean of the robustness values with minimum absolute value over the 20 trials.

For the \CAUR and \CASA approaches, we first generate a 2-way covering array: \[CA(2, 16, (v_1, ..., v_{16}))\] where $v_1, ..., v_{16}$ are each the number of possible discrete values a parameter can take.
The number of possible values is $v_1 = ... = v_5 = 5$ for the colors of the vehicles, $v_6 = ... = v_{10} = 5$ for the models of the vehicles, $v_{11}, v_{12} = 5$ for the Jay-walking Pedestrian shirt and pants colors, $v_{13} = 2$ for the existence (True/False) of fog, $v_{14} = 4$ for the discretized space of the Ego vehicle position, $v_{15} = 4$ for the discretized space of the agent Vehicle 1 position, and $v_{16} = 4$ for the discretized space of the Pedestrian speed.
The size of that covering array is 47, and it covers all possible $2,562$ 2-way combinations, while the number of all possible combinations of all parameters would be $5^{12}\cdot 2 \cdot 4^3$.

Each of the \CAUR and \CASA runs begins with the 47 covering array test cases (i.e., these runs account for 47 of the 200 simulations allotted to each trial).
Then, for the test cases that have the closest robustness value to their target (robustness values closest to $0$), the search continues over the real-valued space for 153 additional simulations.
Hence, each \CAUR and \CASA trial has a total budget of 200 simulations, which is equal to the simulation budget for each \GUR trial.
Also, for the \CAUR and \CASA test cases, we limit the maximum number of search iterations (simulations) starting from a given discrete case ($\pd_{d_j}$) to 50, 
at which time we restart the search at the \emph{next best} discrete case and perform another search; this repeats until the total simulation budget of 200 is spent.
For \CAUR and \CASA, as we target finding a glancing case (close to zero robustness value), we take the absolute value of the robustness.

Fig. \ref{fig:probability_distribution} shows that, compared to the \GUR method, the \CAUR method does a better job of identifying glancing cases and also has the advantage of guaranteeing that any 2-way combination among the parameters are covered in our search space, while still using an unguided approach to explore the real-valued space.

The \CASA method does better than both the \GUR and \CAUR method in identifying glancing behaviors.
As can be seen in Fig. \ref{fig:probability_distribution}, \CASA, on average, can find the test cases that generate a behavior closer to the target (i.e., close to $0$).
The \CASA approach, while maintaining the coverage guarantees inherited from covering arrays, also applies falsification in the continuous space, and so \CASA is expected to perform better than either of the other two methods. 
The results in Fig. \ref{fig:probability_distribution} confirm this analysis.

We believe that these results indicate that an approach combining covering arrays and optimization-based falsification, such as \CASA, can be useful in testing autonomous driving systems, especially if specific combinations of some discrete parameters that cannot be directly discovered by intuition may generate challenging scenarios for perception systems.

\section{Conclusions} \label{sec:conclusions}

We demonstrated a simulation-based adversarial test generation framework for autonomous vehicles.
The framework works in a closed-loop fashion where the system evolves in time with the feedback cycles of the autonomous vehicle's controller, including its perception system, which is assumed to be based on a deep neural network (DNN).
We demonstrated a new effective way of finding a critical vehicle behavior by using 1) covering arrays to test combinations of discrete parameters and 2) simulated annealing to find corner-cases.

Future work will include using identified counterexamples to retrain and improve the DNN-based perception system.
Also, the framework can be further improved by incorporating not only camera data but also other types of sensors. 
The scene rendering may also be made more realistic by using other scene rendering tools such as those based on state-of-the-art game engines.


\bibliographystyle{eptcs}
\bibliography{perception_vnv_bibliography}

\end{document}